# Retinal blood flow speed quantification at the capillary level using temporal autocorrelation fitting OCTA

Yunchan Hwang,[1] Jungeun Won,[1] Antonio Yaghy,[2]
Hiroyuki Takahashi,[1,2] Jessica M. Girgis,[2] Kenneth Lam,[2]
Siyu Chen,[1] Eric M. Moult,[1] Stefan B. Ploner,[3] Andreas Maier,[3]
Nadia K. Waheed,[2] and James G. Fujimoto[1,*]

[1]*Department of Electrical Engineering and Computer Science, Research Laboratory of Electronics, Massachusetts Institute of Technology, Cambridge, MA 02139, USA*
[2]*New England Eye Center, Tufts Medical Center, Boston, MA 02116, USA*
[3]*Pattern Recognition Lab, Friedrich-Alexander-Universität Erlangen-Nürnberg, Erlangen, Germany*
*\*jgfuji@mit.edu*

**Abstract:** Optical coherence tomography angiography (OCTA) can visualize vasculature structures, but provides limited information about the blood flow speeds. Here, we present a second generation variable interscan time analysis (VISTA) OCTA, which evaluates a quantitative surrogate marker for blood flow speed in vasculature. At the capillary level, spatially compiled OCTA and a simple temporal autocorrelation model, $\rho(\tau) = \exp(-\alpha\tau)$, were used to evaluate a temporal autocorrelation decay constant, $\alpha$, as the blood flow speed marker. A 600 kHz A-scan rate swept-source provides short interscan time OCTA and fine A-scan spacing acquisition, while maintaining multi mm$^2$ field of views for human retinal imaging. We demonstrate the cardiac pulsatility and repeatability of $\alpha$ measured with VISTA. We show different $\alpha$ for different retinal capillary plexuses in healthy eyes and present representative VISTA OCTA of eyes with diabetic retinopathy.

## 1. Introduction

The retina has specialized inner retinal and choroidal vasculature as well as dynamic retinal blood flow regulation to meet its high metabolic demand [1, 2]. In retinal diseases, abnormal expression of cytokines and angiogenic factors can disrupt the retinal vasculature, which can lead to macular edema and neovascularization [3-6]. Retinal blood flow speeds have been characterized to study both normal flow regulation and flow alterations associated with pathologies [7-10]. Blood flow speed measurements in capillaries, where oxygen and metabolic substrate exchange occurs, may provide valuable and unique biomarkers to facilitate understanding of normal physiology and pathologies in retina.

Adaptive optics (AO) provides exceptional transverse resolution for in vivo retinal imaging and allows direct measurements of capillary blood flow speeds [11]. AO has been used to study cardiac pulsatility, range of speed and response to stimulation in capillaries and has deepened our understanding of hemodynamics in retinal vasculature [12-16]. However, AO provides limited field of views (FOVs) and requires a montage to acquire multi mm$^2$ FOV, which can be time-consuming. Doppler optical coherence tomography (OCT) is another imaging modality that directly measures blood flow speeds [8, 17]. However, Doppler OCT cannot be used when blood cell velocity is nearly orthogonal to the beam, as in retinal capillaries. OCT with repeated A-scans (M-mode) has been used to characterize flow speeds using temporal autocorrelation, by acquiring >100 A-scan repeats at each position with a temporal resolution of <100 μs [18-21]. Large numbers of data points were fitted to various temporal autocorrelation models to determine flow speeds. However, OCT protocols with a high number of repeats are not suitable for multi mm$^2$ FOV human retinal imaging in a clinical setting.

OCT angiography (OCTA) is a functional extension of OCT, and both OCT and OCTA are standard, commercially available imaging modalities in ophthalmology [22-25]. OCTA generates contrast from time-varying OCT signals, by acquiring two or more repeated B-scans separated by a given time interval (interscan time). Since OCTA is generated from OCT images which have intrinsic depth resolution, it can visualize individual retinal plexuses with great detail [26]. OCTA can detect capillary non-perfusion, neovascularization and microaneurysms (MAs) [27-29]. OCTA derived biomarkers, such as vessel density, non-perfusion area, foveal avascular zone area, have been extensively studied to characterize retinal pathologies using both commercial devices and research prototypes [30-33]. However, OCTA in its traditional form provides only structural information on vasculature and does not evaluate blood flow speeds.

Increases in OCT A-scan rates enabled extensions of OCTA for characterizing retinal blood flow speeds. The OCTA signal saturates when longer interscan times are used and blood flow speed governs how fast the saturation occurs [20]. By comparing OCTA signals measured with two different interscan times, acquired with >2 B-scan repeats, our group previously developed variable interscan time-analysis (VISTA) OCTA using a 400 kHz A-scan rate vertical cavity surface emitting laser (VCSEL) light source [34]. Relative blood flow speeds were mapped to a false color image using the ratio of en face OCTA signals measured with two different interscan times [35]. Using VISTA OCTA, we qualitatively characterized flow variations associated with age-related macular degeneration (AMD), diabetic retinopathy (DR) and polypoidal choroidal vasculopathy (PCV) [34, 36-39].

Here, we introduce a next generation VISTA algorithm which evaluates temporal autocorrelation decay constant in human retinal vasculature at the individual capillary segment level. Higher A-scan rate (600 kHz) swept-source OCT (SS-OCT) enables multiple B-scan repeats with short interscan times and fine A-scan spacing, providing OCTA measurements at multiple interscan times and resolving retinal capillaries. We evaluate the temporal autocorrelation decay constant, $\alpha$, by fitting the OCTA measurements to a simple temporal autocorrelation decay model $\rho(\tau) = \exp(-\alpha\tau)$. We propose $\alpha$ (ms$^{-1}$) as a quantitative surrogate marker for blood flow speeds, assuming that higher blood flow speed will lead to faster autocorrelation decay and higher $\alpha$. Instead of using >100 repeated A-scans, we computationally identify retinal capillary segments and compile OCTA measurements across voxels comprising each capillary segment. Here, we show that $\alpha$ measured from VISTA OCTA demonstrates pulsatility consistent with previous AO studies of capillary blood flow and propose a pulsatility compensation algorithm. We recruited healthy subjects and patients with diabetic retinopathy. We evaluated repeatability of VISTA $\alpha$ measurements at multiple spatial levels as well as consistency of $\alpha$ measurements from OCTA protocols with different interscan times and A-scan spacings. We also develop VISTA methods for choroidal vasculature, whose structure is distinct from inner retinal vasculature. We demonstrate that VISTA OCTA can resolve different blood flow speeds for different retinal capillary plexuses in healthy subjects and flow alterations in diabetic retinopathy patients.

## 2. Methods

### 2.1. System description and imaging acquisition

#### 2.1.1 SS-OCT system

A retinal SS-OCT prototype instrument operating at 600 kHz A-scan rate was developed using a custom micro-electromechanical systems VCSEL (MEMS-VCSEL, Praevium Research) with a 1050 nm center wavelength and ~100 nm sweep bandwidth. The frequency sweep was linearized in time to improve SNR over the wavenumber span and more effectively utilize digitizer bandwidth. The OCT signal and a Mach Zehnder interferometer (MZI) calibration signal were sampled simultaneously using two balanced photodetectors (PDB481C-AC, Thorlabs) and a 12-bit digitizer (ATS9373, AlazarTech) at 2 Gs/s. The MZI signal was used to resample corresponding OCT signal to compensate sweep-to-sweep variation of the VCSEL. The incident power at the pupil was <5 mW, consistent with American National Standards

Institute ocular exposure guidelines. The 1/e² beam diameter at pupil was 1.4 mm. A dual axis galvanometer (6210H, Cambridge Technology) was used to scan the OCT beam on the retina. Refractive error of the eye was corrected using an electrically focus tunable liquid lens (EL-10-30-C-NIR-LD-MV, Optotune) placed before the galvanometer, without translating the OCT optics. The focus tunable liquid lens was controlled by software provided by the manufacturer (Lens Driver Controller, Optotune).

2.1.2 Human imaging and OCTA protocols

Healthy normal subjects and diabetic retinopathy patients were imaged at New England Eye Center of Tufts Medical Center (Boston, MA). Only healthy normal subjects were imaged at Massachusetts Institute of Technology (Cambridge, MA). The study was approved by Tufts Medical Center Institutional Review Boards and Massachusetts Institute of Technology Committee on the Use of Humans as Experimental Subjects. Written informed consent was obtained before imaging. Two different scan protocols were used. Parameters of the protocols are shown in Table 1. The fundamental interscan time refers to the time interval between sequentially repeated B-scans.

**Table 1. OCTA protocols**

| FOV | Fundamental interscan time | A-scan spacing | B-scan repeats | Scanner duty cycle | A-scans / B-scan | B-scans | Acquisition time |
|---|---|---|---|---|---|---|---|
| [a]3 mm × 3 mm | 1 ms | 6.7 μm | 8 | 0.75 | 450 | 450 | 3.60 s |
| 5 mm × 5 mm | 1.25 ms | 8.8 μm | 5 | 0.76 | 570 | 570 | 3.57 s |

[a]Data presented in Fig. 5 was acquired with a slightly modified 3 mm × 3 mm protocol: 0.83 ms interscan time, 7.5 μm A-scan spacing, 10 B-scan repeats, 0.8 duty cycle, 400 A-scans /B-scan, 400 B-scans, 3.33 s acquisition time.

*2.2 OCTA processing and segmentation*

2.2.1 OCTA acquisition

$N$ repeated B-scans having a fundamental interscan time $\Delta t$ can provide OCTA with $N$-1 different effective interscan times from $\Delta t$ to $(N-1)\Delta t$. Unnormalized OCTA and normalized OCTA at $M\Delta t$ interscan time were generated from OCT signals using the following formula.

$$\text{OCTA}_{\text{unnormalized}}(M\Delta t) = \frac{1}{L(N-M)} \sum_{i=1}^{L} \sum_{j=1}^{N-M} |S^i_{j+M} - S^i_j|$$

$$\text{OCTA}_{\text{normalized}}(M\Delta t) = \frac{1}{L(N-M)} \sum_{i=1}^{L} \sum_{j=1}^{N-M} \frac{(S^i_{j+M} - S^i_j)^2}{(S^i_{j+M})^2 + (S^i_j)^2}$$

(1)

$S^i_j$ is $j^{\text{th}}$ repeated measurement of OCT signal measured with $i^{\text{th}}$ split spectrum [40]. $L$ is the number of split spectra. Here we used 3 Gaussian spectral bands in order to better match axial and transverse resolution and reduce noise, with the adjacent Gaussian bands crossing each other at 64% of the peaks.

Eye movement during imaging was corrected by registering the unnormalized OCTA B-scan from each position to the unnormalized OCTA B-scan at the previous position using rigid translation [41]. This inter B-scan registration significantly improved connectivity of retinal vasculature. The inter B-scan registered OCTA and OCT data was then segmented.

2.2.2 Retinal layer segmentation

The retinal pigment epithelium (RPE) was segmented for each B-scan using the structural OCT B-scan. Then, the RPE location was smoothed over the volume using 2D local regression smoothing (MATLAB fit function with lowess option, with 'Span' corresponding to 500 μm × 500 μm). The OCT and OCTA volumes were flattened with respect to the RPE. Then, three additional retinal layers were segmented: the internal limiting membrane (ILM), posterior of retinal nerve fiber layer (RNFL) and center of inner nuclear layer (INL). Simple algorithms using thresholding and peak detection were used to generate preliminary segmentation of the three layers. The segmentation results were then manually inspected and corrected over the volume. During the manual correction phase, the reader provided a rough guidance on the layer location in the volume, and the local maximum of "the feature" near the given guidance was detected as the layer. The features for identifying the three layers were intensity increase along axial direction for ILM, intensity decrease along axial direction for posterior of RNFL, and low intensity for center of INL. Note that the most hypo-reflective location in the INL was approximated to be center of INL. The manually corrected segmentation was smoothed by 2D local regression smoothing (MATLAB fit function with lowess option). The 'Span' option of ILM and center of INL corresponded to 150 μm × 150 μm and 300 μm × 300 μm, respectively. The thickness of the RNFL was smoothed with the 'Span' corresponding to 1000 μm × 1000 μm. The RNFL posterior was located by offsetting the smoothed RNFL thickness from the smoothed ILM. The three retinal layer segmentation was done using the RPE-flattened structural OCT volume, Gaussian filtered with a sigma of 1 pixel axially (2.7 μm in tissue) and 3 × 3 pixels transversely.

Accurate and consistent segmentation of the RPE is crucial for choroidal vasculature segmentation. In order to fine tune the RPE segmentation, the posterior local intensity peak in Bruch's membrane-RPE complex was detected from the RPE-flattened structural OCT volume, which was Gaussian filtered in the transverse direction (sigma 3 × 3 pixels). The resulting fine RPE segmentation was median filtered with a window size of 200 μm × 200 μm and used as the reference for the choroidal vasculature segmentation shown in Fig. 5.

### 2.2.3 Vasculature segmentation

The three segmented inner retinal layers divided the four inner retinal plexuses into three categories: RNFL plexus (RNFLP), superficial capillary plexus + intermediate capillary plexus (SCP + ICP) and deep capillary plexus (DCP). The RNFLP region was between 19 μm posterior to the ILM and posterior to the RNFL. The posterior shift from the ILM excluded regions with specular reflections and thin RNFL, which are reported to be avascular [42]. The SCP + ICP region was between the posterior of the RNFL and center of the INL. The DCP region was between the center of the INL and 80 μm anterior to the RPE.

### *2.3 Temporal autocorrelation fitting*

### 2.3.1 OCTA and temporal autocorrelation decay

The temporal autocorrelation of OCT and OCTA, and their relationship to blood flow dynamics have been studied extensively [18-21]. Here we provide a brief overview of relationship between normalized OCTA and temporal autocorrelation decay. OCT signal from a voxel at $x$ at time $t$ can be separated into static and dynamic components, $S(x, t) = S_{static}(x) + S_{dynamic}(x, t)$, where $E[S_{dynamic}] = 0$. In the retinal vasculature, $S_{dynamic}$ mainly arises from moving red blood cells. Treating $S(x, t)$ as a wide-sense stationary process over a time window that is significantly shorter than the cardiac cycle, the expected value of the squared difference of OCT acquired with interscan time $\tau$ is $E[|S(x, t + \tau)-S(x, t)|^2] = 2(Var[S_{dynamic}] - R_{dynamic}(\tau))$, where $R_{dynamic}(\tau)$ is the autocorrelation function of $S_{dynamic}(x, t)$. Note $Var[S_{dynamic}]$ equals $R_{dynamic}(\tau = 0)$. The expected value of the square of the OCT signal is $E[|S(x, t)|^2] = |S_{static}|^2 + Var[S_{dynamic}]$. We define $A(\tau)$ as the scaled ratio of these two quantities [Eq.(2)] [43].

$$A(\tau) = \frac{1}{2} \frac{\mathrm{E}[|S(x,t+\tau)-S(x,t)|^2]}{\mathrm{E}[|S(x,t)|^2]}$$
$$= \frac{\mathrm{Var}[S_{\text{dynamic}}]}{|S_{\text{static}}|^2 + \mathrm{Var}[S_{\text{dynamic}}]} (1 - \frac{R_{\text{dynamic}}(\tau)}{R_{\text{dynamic}}(\tau=0)}) = \beta(1-\rho(\tau)) \tag{2}$$

$\beta = \mathrm{Var}[S_{\text{dynamic}}]/(|S_{\text{static}}|^2 + \mathrm{Var}[S_{\text{dynamic}}])$ models the ratio of dynamic versus static contribution to the signal, and $\rho(\tau) = R_{\text{dynamic}}(\tau)/R_{\text{dynamic}}(\tau = 0)$ models the normalized autocorrelation (autocorrelation coefficient) decay of the dynamic signal. The portion of the dynamic signal contribution is modeled by $\beta$ and not $\rho(\tau)$. Blood flow velocity will govern how fast the autocorrelation $\rho(\tau)$ decays. Note that the OCTA$_{\text{normalized}}$ formula in Eq.(1) is sampling the ratio of squared difference of the OCT signal to the sum of square of the OCT signal at two time points. Therefore, OCTA$_{\text{normalized}}$ can be treated as sampling of $A(\tau)$ at discrete interscan times $\tau = \Delta t, 2\Delta t, \ldots (N-1)\Delta t$, in which case $\beta$ is saturated value of the OCTA$_{\text{normalized}}$.

Here, we adopted a simple model, $\rho(\tau) = \exp(-\alpha\tau)$, where $\alpha$ is a temporal autocorrelation decay constant. We assume a monotonic relationship between $\alpha$ and blood flow speeds in retinal capillaries (Discussion 4.2), and propose $\alpha$ as a blood flow speed surrogate marker.

### 2.3.2 Using spatial compilation to increase sampling and characterize temporal autocorrelation decay

The stochastic nature of blood flow means that a large number of samples are necessary for robust temporal autocorrelation decay characterization. In M-mode acquisitions, the large sampling is achieved with >100 repeated A-scans and temporal resolution of <100 μs. However, OCT protocols with >100 repeated scans require long imaging times, which are not feasible for clinical retinal imaging. Using OCTA protocols with a significantly fewer number of repeats (<10) compared with M-mode, we obtain large number of samples by spatial compilation of OCTA measurements over a spatial domain where blood flow speeds are expected to be uniform.

For inner retinal vasculature, AO studies have shown that the blood flow speeds in capillaries can change significantly (>2×) at bifurcation points [14, 15], making individual capillary segments the largest spatial domain where uniform blood flow speeds can be assumed. Therefore, we spatially compile OCTA measurements over each capillary segment and evaluate a temporal autocorrelation decay constant for individual capillary segments. Note that alternate spatial domains can be adopted to study different vasculatures such as choriocapillaris / choroidal vasculature or focal vascular lesions.

### *2.4 VISTA pipeline for inner retinal vasculature*

### 2.4.1 Voxel level capillary identification

In order to spatially compile the OCTA measurements for each capillary segment, voxel locations for every individual capillary segment need to be identified. Voxel level capillary identification was achieved by pixel level capillary identification and projection of the capillary ID on the 3D vessel mask (Fig. 1A). First, the Hessian based vesselness response proposed by Jerman et al. was applied on the en face OCTA [44]. The global Otsu threshold from the logarithm of the vesselness response was used to produce a 2D vessel mask. The 2D vessel mask was skeletonized and turned into a graph, where the links represent center lines of the vessel and the nodes represent bifurcation points [45]. All 2D vessel pixels of the en face OCTA were given a corresponding vessel ID, visualized by different colors of segments in Fig. 1A (middle). The vessel ID of each pixel was projected on the corresponding 3D vessel mask, giving each voxel the vessel ID. Acquisition of the 3D vessel mask is explained in (2.5).

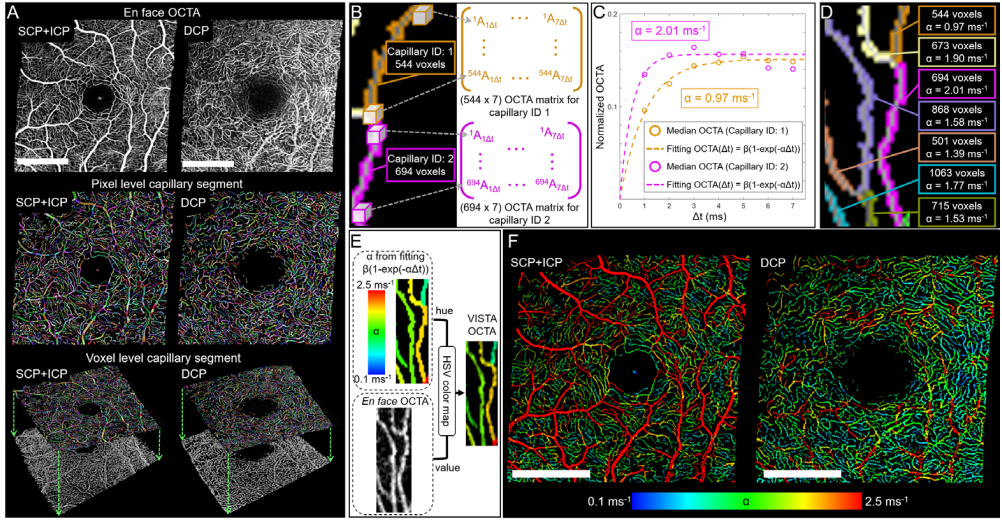

Fig. 1. VISTA pipeline for temporal autocorrelation decay ($\alpha$) evaluation at the capillary level. (A) Voxel level capillary identification. En face OCTA (top) is transformed to a 2D graph (middle). The capillary ID from the 2D graph is projected on a 3D vessel mask (bottom), and is implemented separately for SCP + ICP and DCP. (B) Spatial compilation of normalized OCTA. The spatially compiled OCTA of each capillary can be represented with an OCTA matrix. (C) Temporal autocorrelation decay ($\alpha$) evaluation of two capillaries shown in (B). (D) Number of voxels and corresponding $\alpha$ in the capillaries. (E) Visualization scheme using an HSV colormap to generate VISTA OCTA. (F) VISTA OCTA of SCP + ICP and DCP. Scale bars: 1 mm

### 2.4.2 Spatial compilation of OCTA measurements and OCTA matrix

Voxel level capillary identification enabled compilation of normalized OCTA measurements for each capillary segment. Compiled normalized OCTA measurements of a capillary segment, comprised of $M$ voxels imaged with $N$ B-scan repeats, can be represented with an $M \times (N-1)$ OCTA matrix (Fig. 1B), since $N$ B-scan repeats provide OCTA at $(N-1)$ different interscan times. Medians of the normalized OCTA matrix columns, corresponding to $(N-1)$ interscan time points, were fitted to $OCTA_{normalized}(\tau) = \beta (1 - \exp(-\alpha\tau))$ to evaluate the temporal autocorrelation decay constant, $\alpha$, of the capillary (Fig. 1C). Figure 1D shows the number of voxels comprising the capillary segments and their corresponding $\alpha$.

### 2.4.3 Visualization

To visualize the blood flow speed surrogate marker $\alpha$ in en face OCTA, an HSV colormap (Hue, Saturation, Value) was used, following our previous VISTA implementation (Fig. 1E) [35]. The median filtered (window size: 5 × 5 pixels) $\alpha$ between [0.1 ms$^{-1}$, 2.5 ms$^{-1}$] was linearly mapped to hue between [0.67, 0], which corresponds to [blue, red]. En face OCTA from unnormalized OCTA was used for value in HSV. Saturation was set 1. For visualization purposes, the square root of en face OCTA was used for inner retinal vasculature, due to the high dynamic range of unnormalized OCTA. Note that the hue was determined by normalized OCTA (and its saturation characteristics) while the value was determined by unnormalized OCTA.

### *2.5 3D vessel mask for inner retinal vasculature*

The 3D vessel mask was acquired using Optimally Oriented Flux (OOF) [46]. Given a radius $r$, the OOF response of a voxel at $x$ are three eigenvalues ($\lambda_1, \lambda_2, \lambda_3$, where $|\lambda_1| \geq |\lambda_2| \geq |\lambda_3|$) that quantify the image intensity gradient across the surface of the sphere of radius $r$ along three orthogonal vectors. For example, voxels comprising a planar structure (assuming darker background) will have one large negative eigenvalue with its vector orthogonal to the plane, and two small eigenvalues with two other vectors parallel to the plane, if the radius is comparable to a half of the plane thickness. Voxels comprising a tubular structure will have two large negative eigenvalues with vectors orthogonal to the tube axis and one small eigenvalue with a vector parallel to the tube axis. We used the OOF response proposed by Law and Chung [46], with unnormalized OCTA as input.

$$M(x,r) = \begin{cases} \sqrt{|\lambda_1(x,r)\lambda_2(x,r)|}, & \text{if } \lambda_1, \lambda_2 \leq 0 \\ 0, & \text{otherwise} \end{cases} \quad (3)$$

The radius of the OOF response was set to twice the A-scan spacing. This choice of radius was made to avoid identifying noise as vessels. Note that diameters of human retinal capillaries are smaller than the expected transverse point spread function (PSF) of OCT. The Otsu threshold was applied on the logarithm of the OOF volume as the global threshold to produce a 3D vessel mask.

Figure 2 shows the OOF response of the unnormalized OCTA, with depth color encoded and a vessel mask of the DCP volume. The OOF response suppresses the OCTA projection artifacts because the shape of projection artifacts is planar, making voxels corresponding to projection artifacts have only one large negative eigenvalue, while top voxels corresponding to real vessels have two large negative eigenvalues (Fig. 2A arrows). Thick superficial vessels are mostly removed in the DCP mask, shown in the Fig. 2C. Cross sectional views of the OOF response show three distinct plexuses (SCP, ICP and DCP), with axially oriented, short vessels interconnecting the plexuses (Fig. 2D arrows).

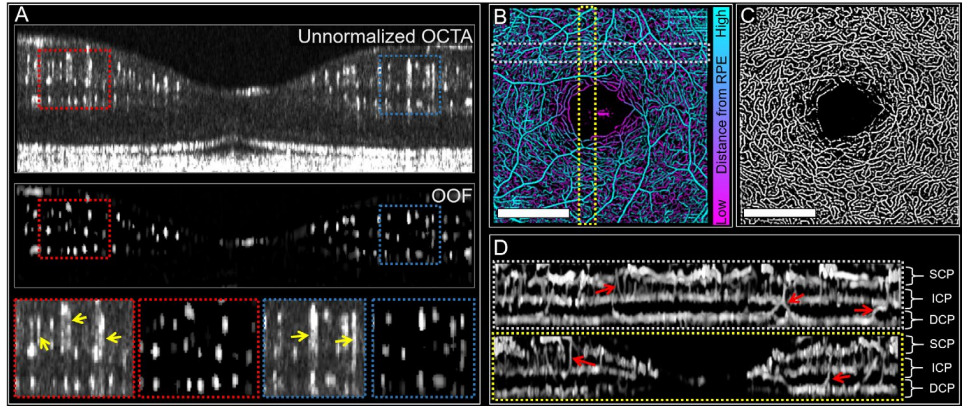

Fig. 2. Optimally Oriented Flux (OOF) for a 3D vessel mask. (A) Unnormalized OCTA B-scan (top) and corresponding OOF response (middle). Enlarged views (bottom) show OOF suppresses projection artifacts in unnormalized OCTA (yellow arrows). (B) En face, depth color-coded OOF. (C) DCP 3D vessel mask from the OOF in (B). (D) Cross sectional views of OOF response. Axially oriented interconnecting vessels are marked with red arrows. Scale bars: 1 mm

*2.6 VISTA in choroidal vasculature*

Since the choriocapillaris and choroid have complex vascular networks that span from lobular structures to thick vessels, choroidal vasculature does not have a single distinct spatial domain, such as the vessel segment in inner retinal vasculature, where we can assume uniform blood flow speed and compile OCTA measurements for evaluating $\alpha$. Therefore, we adopted a rectangular cuboid (53 μm × 53 μm in transverse plane and 8 μm in axial direction) as the spatial domain for OCTA measurements compilation for choroidal vasculature. Note that a different size of cuboid or other vascular structures can be used for the spatial domain if a specific layer of the choroidal vasculature is of interest.

*2.7 Pulsatility and pulsatility compensation in VISTA*

Measuring blood flow speeds or a surrogate marker over a several second total image acquisition time will be subject to pulsatile flow variation during the acquisition. In OCTA acquisition, where multiple B-scan repeats are acquired before the next B-scan, the pulsatility is encoded mainly in the slow scan direction, along the B-scan indices. While the pulsatility itself can be an important biomarker, it is desirable to develop techniques which compensate pulsatility in order to simplify studies with blood flow speed quantification or visualization. For example, pulsatile variations in blood flow speed can complicate studies if alterations caused by pathology are comparable to the pulsatile effects.

### 2.7.1 Modeling and compensating cardiac cycle pulsatility

Pulsatility of the temporal autocorrelation decay constant $\alpha$ at the capillary segment level can be modeled by introducing a time varying scale factor.

$$\alpha^{(i)}(t) = \alpha_0^{(i)}(1 + g^{(i)}(t)), \text{ with } \overline{g^{(i)}(t)} = 0 \tag{4}$$

$\alpha_0^{(i)}$ is the time-averaged temporal autocorrelation decay constant of a capillary segment (i). The cardiac cycle pulsatility is modeled with a pulsatility scale factor $g^{(i)}(t)$. A pulsatility compensation scheme aims to retrieve $\hat{\alpha}_0^{(i)} = \alpha^{(i)} / (1 + \hat{g}^{(i)})$, where $\alpha^{(i)}$ is the temporal autocorrelation decay constant evaluated without pulsatility compensation and $\hat{g}^{(i)}$ is the estimated pulsatility scale factor at the time the capillary was imaged.

### 2.7.2 Pulsatility in VISTA measurements

To characterize the pulsatility of $\alpha$ measured with VISTA, the same B-scan location across the fovea of a healthy subject was repeatedly imaged for 2 seconds with 1 ms interscan time (2000 B-scans of 3 mm length with 450 A-scans/B-scan). This data is essentially an OCT volume which encodes time along one axis (marked yellow in Fig. 3A). Eye movement was corrected using unnormalized OCTA, therefore cross sections of vessels appear as tubular structures aligned to the time axis. OOF was applied to acquire a 3D vessel mask. Roughly 50 capillary cross sections were selected and each cross section was tracked along the time axis, shown as different color tubes in Fig. 3A. With a 100 ms time window, OCTA at interscan times from 1 ms to 7 ms were compiled and fitted to OCTA$_{normalized}(\tau) = \beta (1 - \exp(-\alpha\tau))$. By sliding the 100 ms time window along the time axis, we evaluated the time evolution of $\alpha(t)$ of each capillary and corresponding $g(t)$. Figure 3B shows the time evolutions of the pulsatility scale factors $g(t)$ of the segmented capillaries. The pulsatility scale factors for the capillaries show a certain degree of synchronization, but have both phase and magnitude heterogeneity. In general, the pulsatility factors have fast rise followed by slow recovery, similar to the pulsatile modulation of blood flow speeds in retinal capillaries reported in AO studies [13-15].

### 2.7.3 Representative pulsatility from compiling all vessel voxels

Here, we propose a method to extract representative pulsatility, which can be later extended to pulsatility compensation in VISTA OCTA protocols with isotropic FOVs. First, we compiled the OCTA measurements of all vessel voxels in the 100 ms time window, treating all the vessel voxels as one compiled vasculature (Fig. 3A). Then we evaluated the time evolution of $\alpha$ for the compiled vasculature, $\alpha^{compiled}(t)$ by sliding the time window along the time axis. Since the compiled vasculature includes all capillaries, the pulsatility of the compiled vasculature can serve as the representative pulsatility, $g^{rep}(t)$.

$$\alpha^{compiled}(t) = \alpha_0^{compiled}(1 + g^{rep}(t)), \text{ with } \overline{g^{rep}(t)} = 0 \tag{5}$$

The blue curve in Fig. 3B shows the pulsatility scale factor of the compiled vasculature, $g^{rep}(t)$, which shows a fast rise followed by a slow recovery. Note that the magnitude of $g^{rep}(t)$ is smaller than $g(t)$ of individual capillaries. Compiling multiple capillaries which have different phases (i.e. temporal delays) blunts the rise and fall of $g^{rep}(t)$ but preserves the overall pulsatile variation.

We evaluated Pearson's correlation coefficients among $g(t)$ of individual capillaries and $g^{rep}(t)$ to characterize: i) the degree of pulsatility synchronization and ii) how well $g^{rep}(t)$ represents pulsatility of individual capillaries (Fig. 3C and D). The average correlation coefficient of $g(t)$ with other capillaries is shown in Fig. 3D. A few individual capillaries showed little correlation with other capillaries. The average correlation of the $g^{rep}(t)$ with the individual capillaries was 0.589.

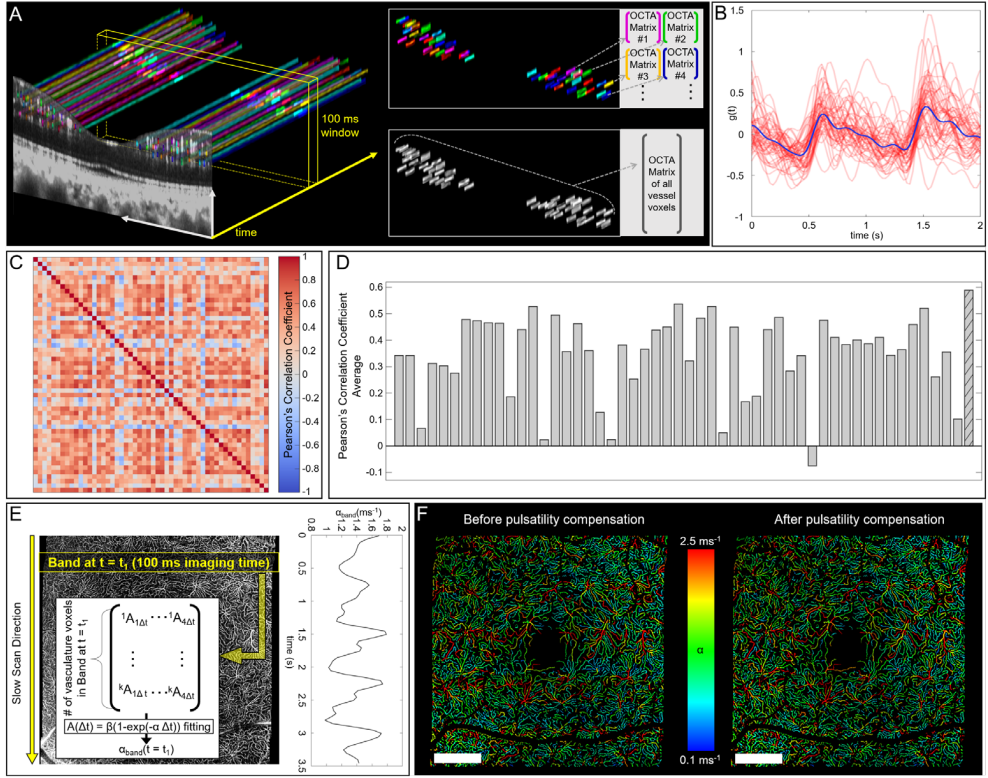

Fig. 3. Pulsatility and pulsatility compensation in VISTA. (A) Repeated B-scans at the same location over time. Compilation of OCTA in the 100 ms time window for $\alpha$ evaluation of the individual capillaries (top right) and the compiled vasculature (bottom right). (B) Pulsatility scale factors, $g(t)$, of individual capillaries (red) and compiled vasculature (blue). (C) Pearson's correlation coefficient matrix between pulsatility scale factors $g(t)$. The right most column is the compiled vasculature. (D) The average of Pearson's correlation coefficient with other capillaries. The right most bar (patterned) is the compiled vasculature. (E) Extracting representative pulsatility from isotropic OCTA protocols. The plot on right shows the changes in $\alpha^{band}$ over slow scan direction, demonstrating clear pulsatility. (F) VISTA OCTA before pulsatility compensation (left) and after pulsatility compensation (right). The plot in (E) is also aligned with the VISTA OCTA images in (F). Scale bars: 1 mm

2.7.4 Pulsatility compensation in isotropic FOV OCTA protocols

In the 2000 repeated B-scan data set (2.7.3), the 100 ms time window was used to compute $g^{rep}(t)$ by compiling all vessel voxels in the time window. Here we propose a general method to extract the representative pulsatility for OCTA protocols with isotropic (or near) FOVs and a pulsatility compensation scheme. In OCTA protocols with isotropic FOVs, the slow scan axis encodes both i) time evolution and ii) spatial sweep in the slow scan direction. If we assume the pulsatility is uniform across the FOV, the slow scan direction encodes essentially only time evolution of the pulsatility. Therefore, we propose to measure α over a "band" that covers 100 ms imaging time, treating all vessel voxels in the band as one compiled vasculature, and slide the band in the slow scan direction (Fig. 3E). Then, the pulsatility of $\alpha^{band}(y_n)$, where $y_n$ is the slow scan index of the band, is the representative pulsatility where $y_n$ encodes the time linearly ($t_{imaging} = y_n N \Delta t$).

$$\alpha^{band}(y_n) = \alpha_0^{band}(1 + g^{rep}(y_n)), \text{ with } \overline{g^{rep}(y_n)} = 0 \tag{6}$$

The pulsatility compensated $\hat{\alpha}_0$ can be estimated by normalization using Eq. (7).

$$\hat{\alpha}_0(x_n, y_n) = \alpha(x_n, y_n) / (1 + g^{rep}(y_n)) \tag{7}$$

$x_n$ is the A-scan index in the B-scan. Fig. 3F shows pulsatility compensation of the DCP. Note that this method only compensates the representative pulsatility and does not account for both temporal and magnitude heterogeneity of the pulsatility among capillaries. On average, we expect our scheme to under-compensate the pulsatility, because compiling all vessel voxels is expected to reduce the magnitude of pulsatility, as shown in Fig. 3B. For choroidal vasculature pulsatility compensation, all voxels in the sliding band were treated as compiled vasculature.

### 2.8 Data acquisition for repeatability and consistency evaluation of VISTA measurements

For repeatability evaluation of the proposed VISTA method, four healthy normal eyes from 4 volunteers (age range from 26 to 40 years old) were imaged with the 3 × 3 mm protocol. The subjects were instructed to sit back from the instrument headrest and chinrest between repeated acquisitions. Subsequently, the subjects were realigned using a mechanical joystick and refocused using the focus tunable liquid lens. For evaluation of the consistency between VISTA measurements from different protocols, one eye (subject #1) was imaged with the 5 × 5 mm protocol after the 3 × 3 mm repeatability protocol image acquisition.

### 3. Results

### 3.1 Comparing SCP + ICP and DCP in healthy subjects

VISTA α measurements in SCP + ICP and DCP from 4 healthy subjects are shown in Fig. 4A. $\hat{\alpha}_0^{SCP+ICP}$ and $\hat{\alpha}_0^{DCP}$ were calculated by averaging $\hat{\alpha}_0(x_n, y_n)$ at vascular skeletons within 1.5 mm eccentricity at each vascular plexus. The SCP + ICP showed faster autocorrelation decay constant than the DCP, suggesting SCP + ICP has faster blood flow speeds than DCP within the eccentricity of 1.5 mm. The ratios of $\hat{\alpha}_0^{DCP}$ to $\hat{\alpha}_0^{SCP+ICP}$ were ~0.7 and were comparable among the healthy subjects (Fig. 4B). The subjects were imaged with the 3 × 3 mm protocol for 4 times (2.8).

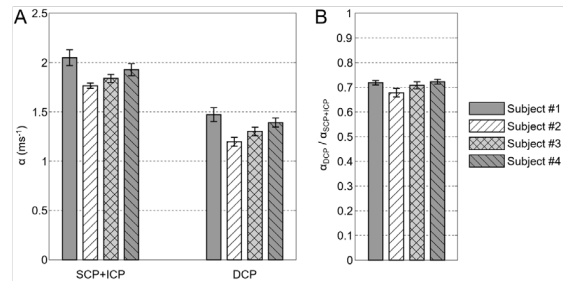

Fig. 4. VISTA α measurements at different plexuses in healthy subjects (age ranging from 26-40 years old).

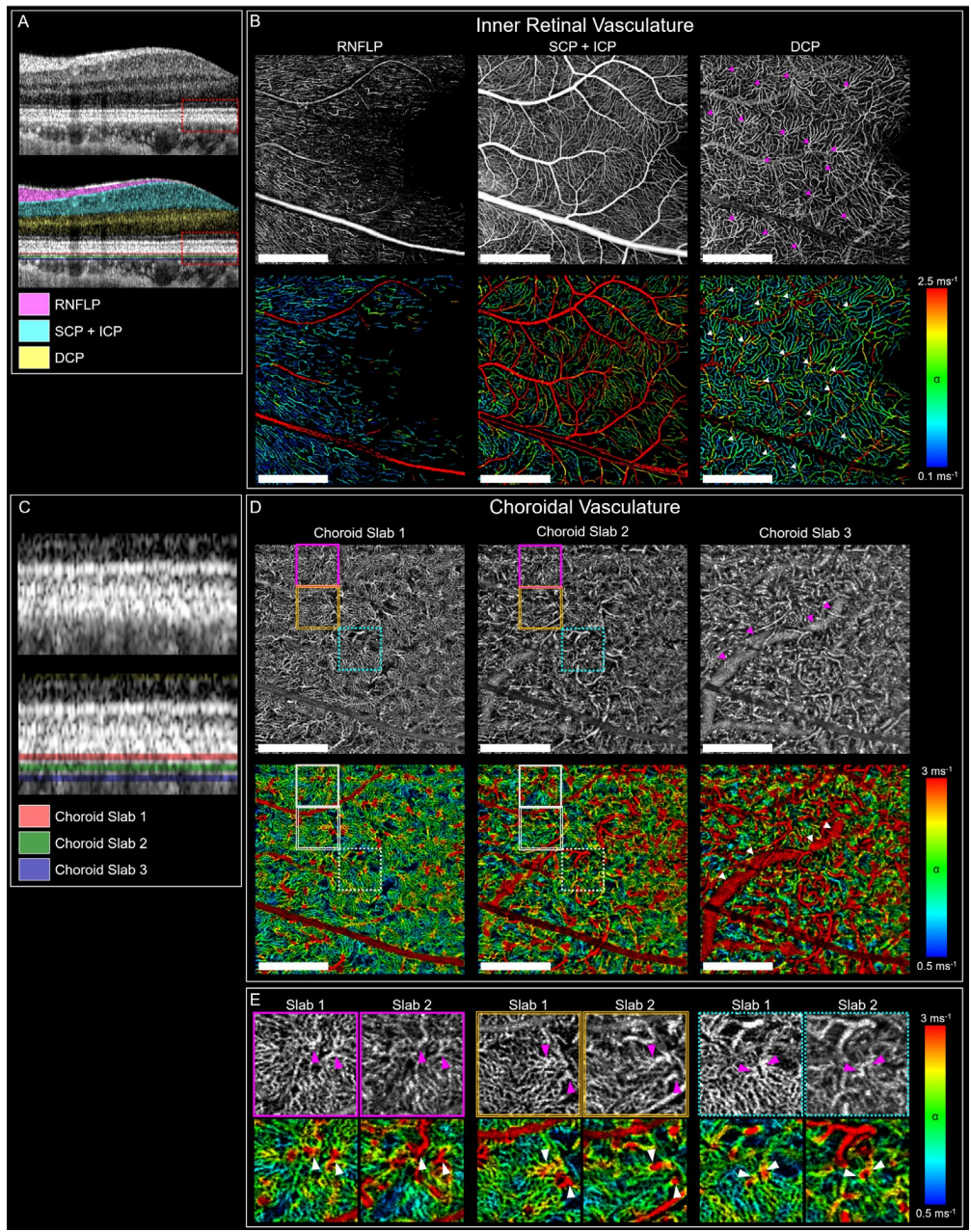

Fig. 5. VISTA OCTA of inner retinal and choroidal vasculature of a healthy subject. (A, C) Segmentation of RNFLP, SCP + ICP, DCP, and Choroidal slabs 1, 2 and 3. (B) En face OCTA (top) and VISTA OCTA (bottom) of RNFLP, SCP + ICP and DCP. Centers of distinct vortex-like and fern-like patterns of DCP are marked with arrowheads, which often show high $\alpha$ in VISTA OCTA. (D) En face OCTA (top) and VISTA OCTA (bottom) of Choroidal slabs 1, 2, and 3. Three boxes in Choroidal slab 1 mark choriocapillaris lobules. The same regions are marked in Choroidal slab 2 to compare vascular structure and flow speeds posterior to the lobules. A thick vessel in Choroidal slab 3 is marked with arrowheads. (E) Enlarged en face OCTA (top) and VISTA OCTA (bottom) of the three boxes in (D). The centers of the choriocapillaris lobules in Choroidal slab 1 are marked with arrowheads, which show high $\alpha$ (left). Thicker choroidal vessels of ~50 µm diameters are posterior to the centers of lobules (right). Scale bars: 1 mm.

## 3.2 Retinal and choroidal vasculature with VISTA

Figure 5 shows VISTA OCTA of 6 different vasculatures from one data set of a healthy subject, imaging the region between optic nerve head (ONH) and fovea. Segmentation of the RNFLP, SCP + ICP and DCP are shown in Fig. 5A. The SCP + ICP has higher $\alpha$ than DCP, and RNFLP has lower $\alpha$ than DCP. Distinct vortex-like and fern-like vascular patterns of DCP are marked with white arrowheads in Fig. 5B. We observed the center of these vascular patterns often have high $\alpha$ in VISTA OCTA (arrowheads). Segmentation of three choriocapillaris / choroidal slabs are shown in Fig. 5C. Slabs 1, 2 and 3 are 13 μm, 27 μm and 40 μm posterior to RPE, respectively. The thickness of the slabs are 8 μm. Slab 1 in en face OCTA shows lobular structures of the choriocapillaris. The centers of several lobules, which have maximal connection to the branches of the lobules, are marked with arrowheads in Fig. 5E. The same locations are marked in Choroidal slab 2 and show thicker choroidal vessels of ~50 μm diameters. VISTA OCTA of Fig. 5E shows the centers of the lobules have higher $\alpha$ compared to their surroundings. We believe this suggests that the center of lobular structures have higher blood flow speeds than the periphery of the lobules. Choroidal slab 3 shows a thicker vessel (~150 μm diameter, arrowheads), which has high $\alpha$ in VISTA OCTA (Fig. 5D).

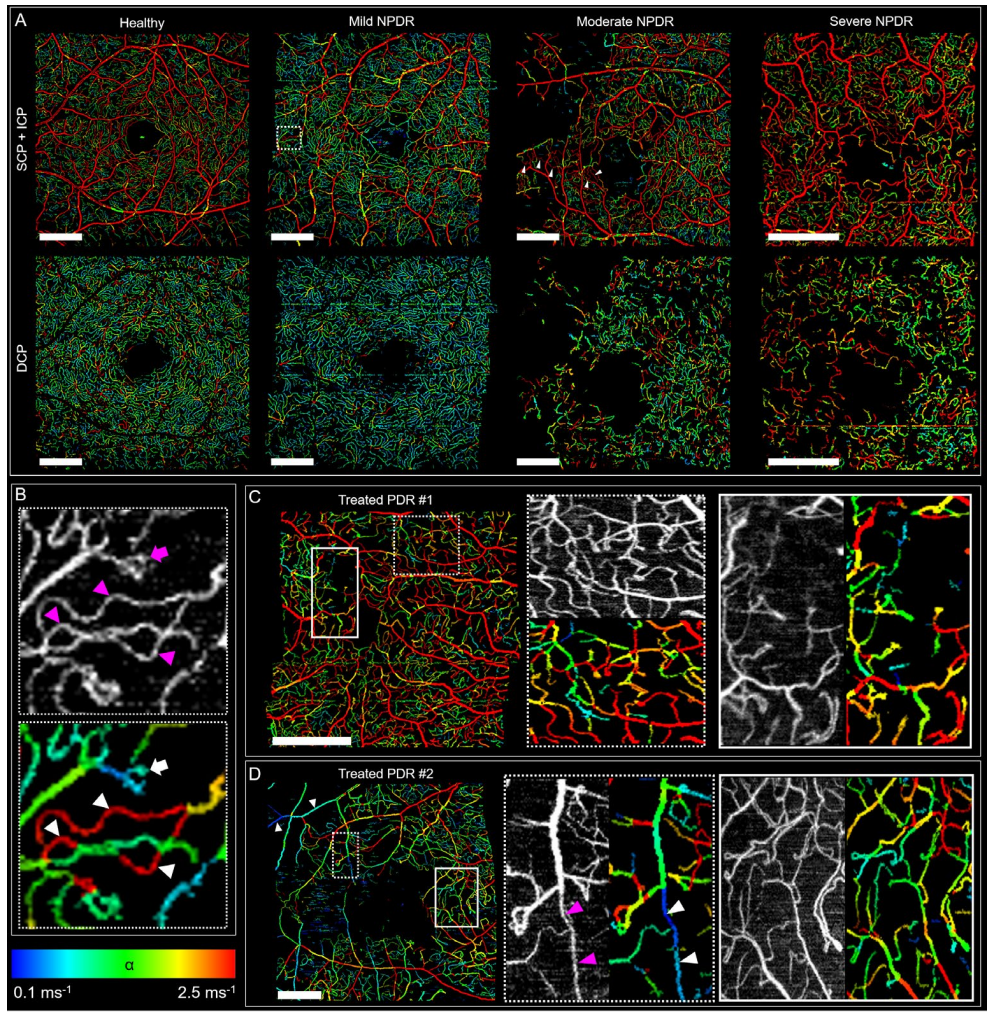

Fig. 6. VISTA OCTA of a healthy eye and eyes with different stages of diabetic retinopathy.
(A) VISTA OCTA of a healthy eye and eyes with mild, moderate, and severe non-proliferative

diabetic retinopathy (NPDR). A dashed box marks the region with low capillary density in the mild NPDR, enlarged in (B). Arrowheads in SCP + ICP of the moderate NPDR mark tortuous capillaries with high $\alpha$. (B) Arrowheads and an arrow mark the tortuous vessels and the focal bulge. (C, D) VISTA OCTA of treated proliferative DR (PDR) with enlarged views on right. (D) Arrowheads mark the thick vessels (not capillaries) that show low $\alpha$. Scale bars: 1 mm.

### 3.3 Diabetic retinopathy with VISTA

Representative VISTA OCTA images of the SCP + ICP and DCP of a healthy eye as well as mild, moderate and severe non-proliferative diabetic retinopathy (NPDR) eyes are shown in Fig. 6A. We observed generally low $\alpha$ in the mild NPDR eye. Figure 6B shows an enlargement of the SCP + ICP of the mild NPDR eye from a region with low capillary density. Note the tortuous vessels (arrowheads) and the focal bulge (arrow) have high and low $\alpha$, respectively. In the moderate NPDR eye, we observed tortuous capillaries which had high $\alpha$ (arrowheads). The severe NPDR eye was anti-VEGF treatment naïve at the time of imaging, and showed high $\alpha$, even in the DCP. VISTA OCTA images of the SCP + ICP of treated proliferative diabetic retinopathy (PDR) eyes are shown in Fig. 6C and D. Note that even a few thick vessels in Fig. 6D have low $\alpha$ (arrowheads). Enlarged views of microvascular lesions such as capillary loops and microaneurysms show that VISTA OCTA provides blood flow speed information that cannot be detected using standard OCTA. We emphasize that en face OCTA and the 'intensity' (value of HSV) of VISTA OCTA images are derived from unnormalized OCTA, and only the 'color' (hue of HSV) is derived from normalized OCTA and its saturation characteristics.

### 3.4 Repeatability measurements in healthy subjects at the capillary segment level

In order to evaluate repeatability at the capillary segment level, >100 individual capillary segments from SCP + ICP and DCP respectively were manually traced and the corresponding capillary segments were identified in 4 repeated volumes from subject #1 (healthy). For each segment, $\hat{\alpha}_0(x_n, y_n)$ along the corresponding vascular skeleton was averaged to represent $\hat{\alpha}_0^{segment}$. Figure 7A shows scatter plots of 4 $\hat{\alpha}_0^{segment}$ measurements. The [mean, median] of coefficient of variance (CV) for individual capillary segment measurements were [0.194, 0.164] for SCP + ICP and [0.160, 0.154] for DCP. This is comparable to the reported coefficient of variance of blood flow speeds in retinal capillaries measured using AO [47]. However, note that VISTA measures the temporal autocorrelation decay constant in the vasculature, not the blood flow speed directly.

### 3.5 Repeatability measurements at regional level

In order to evaluate repeatability at the regional level, 4 sub-regions (nasal, superior, temporal, inferior) from SCP + ICP and DCP were traced and subsequently identified in 4 repeated volumes each from 4 healthy subjects. For each region, $\hat{\alpha}_0(x_n, y_n)$ at vascular skeletons in the region was averaged to represent $\hat{\alpha}_0^{region}$. The sum of 4 sub-regions was used to evaluate $\hat{\alpha}_0^{global}$. The CV of $\hat{\alpha}_0^{region}$ and $\hat{\alpha}_0^{global}$ was below 0.075 and 0.05 respectively for both SCP + ICP and DCP and for all subjects (Fig. 7B). To examine the dependence of repeatability on pulsatility compensation, the CV of $\alpha_0^{global}$ without pulsatility compensation was evaluated. The CV was comparable to the case with pulsatility compensation (Fig. 7B). This is expected, since the overall effect of pulsatility compensation across FOV will converge to unity and normal healthy subjects do not have local vascular lesions.

### 3.6 Consistency between different imaging protocols

Depending on the SS-OCT A-scan rate, the FOV, and galvanometer scan speed, the parameters for VISTA OCTA protocols can be different. Since temporal autocorrelation decay fitting and OOF response are the backbone of the new VISTA algorithm, we expect the key parameters to be i) the fundamental interscan time, ii) the number of B-scan repeats and iii) the A-scan spacing. In order for temporal autocorrelation decay constant $\alpha$ measured from VISTA to be a quantitative surrogate marker for blood flow speeds, $\alpha$ measured from different protocols should be consistent, making different VISTA OCTA scan protocols compatible. We imaged subject #1 with both the 3 × 3 mm and 5 × 5 mm protocol (Table 1). The capillary segments and regions marked for the repeatability evaluation in the 3 × 3 mm volume were identified in the 5 × 5 mm volume (Fig. 7C). Figure 7D plots the $\hat{\alpha}_0^{segment}$, $\hat{\alpha}_0^{sub\text{-}region}$ and $\hat{\alpha}_0^{global}$ measurement from the 3 × 3 mm protocol on *x*-axis and the 5 × 5 mm protocol on *y*-axis. Note that $\hat{\alpha}_0^{region}$ and $\hat{\alpha}_0^{global}$ agree well. To examine the bias (consistent over or under estimation of $\hat{\alpha}_0$) caused by different key parameters, '$y = kx$' was fitted in between the two measurements. Fitting $\hat{\alpha}_0^{segment}$ and $\hat{\alpha}_0^{region}$ from the two protocols resulted '$y = 0.969x$' (dashed line in Fig. 7D) and '$y = 0.979x$' (solid line) respectively, indicating a small bias between the results from the two scan protocols.

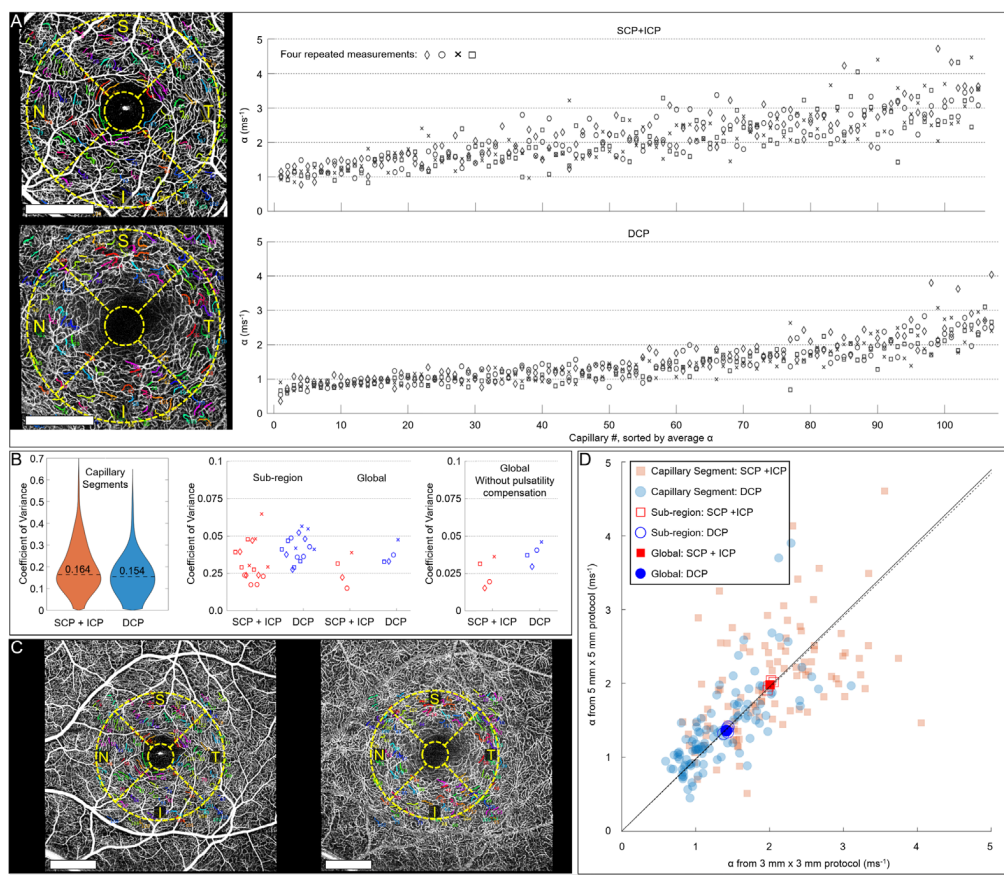

Fig. 7. Repeatability and consistency of VISTA $\alpha$ measurements at different spatial levels. (A) Repeated measurements at the capillary segment level. Color-coded vessels mark the selected capillary segments. Sub-region maps are overlaid on en face OCTA (left) Top row: SCP + ICP. Bottom row: DCP. (B) Coefficient of variance (CV) at different spatial levels. Left: Violin plots of CV of capillary segments. Dashed lines mark medians. Middle and right: Different markers are used for different subjects. Middle: Sub-region includes nasal, superior, temporal and inferior. (C) En face OCTA of 5 mm × 5 mm. Corresponding capillary segments from (A) are color-coded. Left: SCP + ICP. Right: DCP. (D) Comparing $\alpha$ measurements from the 5 mm × 5mm protocol and the 3 mm × 3mm protocol at different spatial levels. Scale bars: 1 mm.

## 4. Discussion

*4.1 Blood flow speed imaging with VISTA*

Figure 4 and 5 show that VISTA can enable studies of blood flow speeds of healthy normal retinal vasculature at the capillary level, while still providing multi mm$^2$ FOV, a scale significantly larger than a few selected vessel segments. To our knowledge, comparing blood flow surrogate markers between the SCP + ICP and DCP has not been reported in humans in vivo. Also, clear visualization of the lobular structure of choriocapillaris in Fig. 5E and with high $α$ near the center of the lobules also suggests that VISTA can be applied to assess choriocapillaris flow deficits which are important markers of disease [48-52]. In pathologies such as DR (Fig. 6), focal vascular lesions can be associated with flow markers. Examples are i) correlations between $α$ of capillary segment and their tortuosity, and ii) comparison of $α$ in/near DR lesions to the average $α$ of the plexus. The regional measurements showed high repeatability and compatibility between different OCTA protocols (Fig. 7). Note that VISTA can quantitatively characterize the blood flow speeds using a single global number, such as $\hat{α}_0^{\text{SCP+ICP}}$ or $\hat{α}_0^{\text{DCP}}$, as well as with the distribution of $α$. The capability of VISTA to provide simple quantitative measurements for an eye, analogous to OCTA vessel density and foveal avascular zone area, simplifies interpretation and can facilitate longitudinal studies on progression or treatment responses.

In addition, we emphasize that the new VISTA protocol also provides standard en face OCTA. This creates a natural connection between future VISTA studies and the existing extensive body of OCTA research. Although a high A-scan rate is required to perform VISTA, smaller FOVs can be assessed using current commercial instruments with 100 to 200 kHz A-scan rates and FOVs can be increased as faster A-scan rates become available.

*4.2 Vascular structures where VISTA can be applied*

Transverse PSF, interscan time and blood flow speeds in the retina can provide a rough estimate of which vasculature structures can be characterized using OCTA temporal autocorrelation fitting. A 1050 nm beam with 1.4 mm 1/e$^2$ diameter, focused with a 17 mm focal length lens assuming low aberration, produces a PSF of ~16 μm 1/e$^2$ beam diameter (~10 μm FWHM) on the retina. Although the optical aberrations of human eyes would make actual transverse PSF on the retina larger, this ideal transverse PSF enables a conservative estimate of which vasculature structures can be characterized using $ρ(τ)$. In healthy normal subjects, blood flow speeds in thick (>40 μm diameter) vessels have been reported to be >~10 mm/sec [53]. With a 1 ms fundamental interscan time, the blood cell displacement between the repeated B-scan will be >~10 μm. Therefore, in transversally oriented thick vessels, $ρ(τ)$ will be nearly fully decayed even at the shortest interscan time. In this case, where OCTA is nearly fully saturated at $1Δt$, the variation in OCTA signal at longer interscan times is likely to be dominated by noise, rather than the actual temporal autocorrelation decay of the vasculature, and fitting OCTA$_{\text{normalized}}(τ)$ = $β$ (1 - exp(-$ατ$)) will yield a high $α$. Conversely, blood flow speeds in retinal capillaries are mostly <3 mm/sec [7, 11, 14]. Therefore, we expect the temporal autocorrelation in retinal capillaries will not be fully decayed at short interscan times, and the decay characteristics can be captured by the VISTA algorithm. The beam diameter at pupil can be decreased to enlarge the transverse PSF at retina, varying the maximum blood flow speed that can be characterized with temporal autocorrelation decay, at the cost of transverse image resolution.

*4.3 Temporal autocorrelation decay model*

Various types of temporal autocorrelation functions, $ρ(τ)$, in OCT have been suggested for assessing blood flow. Wang et al. approximated the amplitude of backscattered light as a sequence of square pulses and modeled autocorrelation function as $ρ(τ) = 1 - α'τ$ when $τ$ is small [19]. The authors experimentally validated a linear relationship between $α'$ and the transverse velocity using intralipid particles. Tokayer et al. demonstrated a linear relationship between normalized OCTA and particle speeds when the interscan time is small, using whole blood flowing in a glass tube [20]. The time range where the linear relationship holds was dependent on the blood flow speeds (i.e. <0.3 ms for 1.5 mm/sec blood flow speed). Note that 1.5 mm/sec is in the range of retinal capillary blood flow speeds and 0.3 ms is significantly smaller than the shortest interscan time we used. (The beam $1/e^2$ diameter was 20 μm in the Tokayer et al. study.) Lee et al. proposed a dynamic light scattering OCT which comprehensively modeled the static and noise component of OCT signal [18]. The authors treated diffusion and translational movement of particles separately and the autocorrelation function included a term proportional to $\exp(-\alpha_1\tau - \alpha_2\tau^2)$. Recently, Nam et al. modeled the autocorrelation as $\rho(\tau) = \exp(-\alpha_1^2\tau^2)$ and measured pulsatility in thick retinal and choroidal vessels in human retina [21].

Adopting a clinically feasible multi mm$^2$ FOV OCTA protocols limited our choice of temporal autocorrelation function. Even the shortest interscan time of ~1 ms was too long for a linear model, following Tokayer et al. [20]. Because the number of temporal points for $\rho(\tau)$ is limited (<10), models with multiple parameters may lead to overfitting. Therefore, we chose $\rho(\tau) = \exp(-\alpha\tau)$ due to its simplicity and ease of interpretation of $\alpha$ as temporal autocorrelation decay constant. The single file flow of blood cells and predominantly fixed direction of motion in retinal capillaries may provide guidance on improving the autocorrelation model $\rho(\tau) = \exp(-\alpha\tau)$. Also, it is possible that retinal capillaries and choriocapillaris have different temporal autocorrelation characteristics, since their hemodynamics are different.

### 4.4 OCT system requirements for VISTA

Assuming that $\rho(\tau) = \exp(-\alpha\tau)$ is a reasonable model for temporal autocorrelation decay in retinal capillaries, the measured $\alpha$ can provide guidance on the fundamental interscan time required for $\alpha$ evaluation using the VISTA algorithm. For example, consider a 3 ms fundamental interscan time using the previous OCT beam size (1.4 mm $1/e^2$ diameter at pupil). At the shortest interscan time, the OCTA signal will already be >99% saturated in the SCP + ICP because the average $\hat{\alpha}_0^{\text{SCP+ICP}}$ is ~2 ms$^{-1}$. We propose using <95% OCTA saturation at the shortest interscan time, or $\exp(-\alpha\Delta t) > 0.05$, as a phenomenological requirement evaluating the autocorrelation decay. This yields $\alpha_{\max}\Delta t_{\text{fundamental}} \approx 3$. Assuming the same beam size at the pupil as the previous example, a fundamental interscan time <1.5 ms will be required to characterize the SCP + ICP. Note that the shortest fundamental interscan time achievable is often determined by the characteristics of the galvanometer scanner.

A key advantage of VISTA protocol is that it also generates standard OCTA images, which have been extensively used to study retinal disease. If the maximum interscan time $(N - 1)\Delta t_{\text{fundamental}}$ in the OCTA protocol is too short, or $\alpha$ is too small, OCTA will have limited contrast, limiting the VISTA protocol from generating OCTA. We propose >50% saturation of OCTA at the longest interscan time, or $\exp[-\alpha(N-1)\Delta t] < 0.5$, as a requirement for sufficient OCTA contrast. This yields $\alpha_{\min}(N - 1)\Delta t_{\text{fundamental}} \approx 0.7$.

Finally, fine A-scan spacing is required in order to resolve capillaries for correct spatial compilation and to achieve a large number of samples. If the A-scan spacing is not fine enough, individual capillaries are less likely to be continuously resolved. We have experimentally determined that a ~10 μm or finer A-scan spacing is required to robustly resolve retinal capillaries.

The fundamental interscan time and A-scan spacing requirement provide a constraint on the A-scan rate and the maximum FOV for the VISTA protocol. For example, assuming a fundamental interscan time of 1.5 ms, 10 μm A-scan spacing, and 75% scanner duty cycle, a ~530 kHz A-scan rate will be required to achieve 6 mm FOV in the fast scan direction. The A-scan rate requirement can be reduced linearly by sacrificing the fast scan FOV. Montaging, anisotropic FOVs, using a smaller beam diameter at the pupil, and improving the scanner duty cycle can also further reduce the A-scan rate requirement.

*4.5 Spatial compilation approaches*

The capillary segment spatial compilation approach was motivated by distinct layered structure of inner retinal vasculature and blood flow speed changes at capillary bifurcation points. However, more general spatial compilation approaches are also possible. For example, the radial peripapillary capillary near the ONH has a radial orientation. Therefore, disc sectors defined by the angle from the ONH can be used for spatial compilation. Different spatial compilation methods may be needed to assess focal vascular lesions. For example, microaneurysms can be inaccurately segmented if the target radius of OOF is smaller than the size of microaneurysms or the 2D vesselness filter does not have high sensitivity to spherical structures. In these cases, different segmentation methods, such as deep-learning, can be used to detect microvascular lesions and determine the appropriate spatial domain for OCTA compilation. However, a chosen spatial domain needs to support the sufficiently large numbers of samples. If the domain does not have enough voxels, the saturation characteristic of the domain may be dominated by noise.

*4.6 Limitations to pulsatility compensation*

If there is eye motion or blinking, the pulsatility compensation may introduce artifacts which will limit OCTA compilation from the sliding band in the slow scan direction (Fig. 3E). Using the sliding band with a longer imaging time window may make evaluation of $\alpha^{band}$ more robust, at the cost of reduced time resolution. Also, our pulsatility compensation scheme assumes the same degree of pulsatility across FOV, which may not hold if there are local lesions such as edema. Cardiac-gated multiple OCTA acquisitions can be used to reconstruct the pulsatile flow variations at all positions in the FOV [17]. However, this requires longer acquisition times and large datasets.

4.7 Potential of VISTA for patient risk stratification

VISTA may offer opportunities for patient risk stratification. For example, neovascularization is a vision impairing condition that can develop secondary to many retinal diseases. The development of neovascularization has been associated with changes in the structural integrity of the retinal vasculature; vascular endothelial growth factor (VEGF) for example, crucial in neovascularization, also increases the permeability of the vasculature [3, 4]. If the change in the vasculature integrity, before the onset of the neovascularization, causes blood flow alterations in capillaries, VISTA blood flow markers might enable identification of high risk patients.

**5. Conclusion**

We developed and demonstrated a second generation VISTA OCTA algorithm that provides a quantitative surrogate marker for blood flow speeds. SS-OCT with a 600 kHz A-scan rate enabled clinically feasible imaging protocols with fine A-scan spacing over multi mm$^2$ FOV and short interscan times (~1 ms), within a total acquisition time of 3.6 s. Compiled normalized OCTA measurements at multiple interscan times were fitted to a temporal autocorrelation decay model to evaluate a temporal autocorrelation decay constant $\alpha$. Fine A-scan spacing facilitated capillary identification at individual capillary segment level for spatial compilation of OCTA. VISTA OCTA measured $\alpha$ showed pulsatility in retinal capillaries consistent with the previous AO measurements, and a pulsatility compensation scheme was demonstrated. We evaluated repeatability of VISTA OCTA and consistency of $\alpha$ measurements from different imaging protocols at multiple spatial levels. We observed blood flow speed differences among retinal vascular plexuses in healthy eyes and blood flow speed alterations in eyes with DR. Finally, OCT system requirements for VISTA and different spatial compilation tactics were discussed. These advances promise to enable clinical studies of blood flow speed alterations in diseases such as diabetes and diabetic retinopathy, providing earlier markers of disease, progression and response to therapy.

## Acknowledgements

This work was supported by the National Institutes of Health R01EY011289 (Bethesda, MD); Beckman-Argyros Award in Vision Research (Irvine, CA); Champalimaud Vision Award (Lisbon, Portugal); Greenberg Prize to End Blindness; Retina Research Foundation (Houston, TX); Topcon Medical Systems (Tokyo, Japan); Massachusetts Lions Eye Research Fund (Belmont, MA); Research to Prevent Blindness (New York, NY). The sponsor or funding organization had no role in the design or conduct of this research.

## Disclosures

EMM: IP related to VISTA-OCTA (P). SBP: IP related to VISTA-OCTA (P). NKW: Topcon (C, F), Complement Therapeutics (C), Olix Pharma (C), Iolyx Pharmaceuticals (C), Hubble (C), Saliogen (C), Syncona (C), AGTC (E), Ocudyne (I), Gyroscope (I), Nidek (F, R), Zeiss (F). JGF: Optovue (I, P), Topcon (F), IP related to VISTA-OCTA (P).

## Data availability

Data underlying the results presented in this paper are not publicly available at this time but may be obtained from the authors upon reasonable request.